\documentclass[doublecol]{epl2} 
\usepackage{amsmath}
\usepackage{mathrsfs}
\usepackage{eucal}
\usepackage{amssymb}
\usepackage{amsfonts}
\usepackage{float}
\usepackage{ragged2e}
\usepackage{hyperref}
\hypersetup{colorlinks,citecolor=blue}
\usepackage{xcolor}
\title{Probing Baryogenesis in $\bm{f(Q)}$ gravity}
\shorttitle{Probing Baryogenesis in ${f(Q)}$ gravity} 

\author{Sai Swagat Mishra\inst{1}\footnote{E-mail: saiswagat009@gmail.com}
\and Aaqid Bhat\inst{1}\footnote{E-mail: aaqid555@gmail.com}
\and P.K. Sahoo\inst{1}\footnote{E-mail: pksahoo@hyderabad.bits-pilani.ac.in}
}
\institute{\inst{1} Department of Mathematics, Birla Institute of Technology and
Science-Pilani,\\ Hyderabad Campus, Hyderabad-500078, India.
}
\abstract{
The origin of matter domination in the Universe is one of the most exciting open puzzles in particle physics and cosmology. Despite many theoretical developments, the actual reason behind baryon-antibaryon asymmetry is still unknown. Our aim here is to examine this phenomenon in the framework of modified gravity theories, which have impressively elucidated the contemporary accelerated expansion of the Universe as well as the early phase. Consequently, this letter sets its sights on the task of constraining a specific variant of modified gravity, namely, $f(Q)$ gravity, in conjunction with gravitational baryogenesis.  The Power Law model and recently proposed DGP-like $f(Q)$ models are considered to find the baryon-to-entropy ratio and compare them with the observed value, that is ${n_B}/s=9.42\times 10^{-11}$. Furthermore, we impose constraints on the additional degrees of freedom introduced by this modified theory of gravity.}

\begin{document}
\maketitle
\section{Introduction }

The observational evidence, as exemplified by phenomena like Big-Bang Nucleosynthesis (BBN) \cite{Burles/2001} and meticulous measurements of the Cosmic Microwave Background (CMB) \cite{Bennett/2003,Spergel/2003}, when coupled with the comprehensive data on the large scale structure of the Universe, unequivocally substantiates a profound revelation: the prevalence of matter vastly outweighs that of its antimatter counterpart. These empirical observations stand as an indisputable testament to the striking asymmetry that exists between matter and antimatter, tracing its origins back to the primordial stages of the Universe, antecedent even to the epoch of Big Bang Nucleosynthesis (BBN). The enigma surrounding the genesis of the baryon asymmetry stands as one of the paramount inquiries in contemporary physics. Delving into the depths of theoretical physics, we encounter a domain known as Baryogenesis, which strives to elucidate the intricate mechanisms that engender the profound disparity between baryons and antibaryons in the nascent Universe.

Within this realm, a multitude of theories emerges, each proposing distinct interactions beyond the confines of the standard model, with the aim of unraveling the origin of this asymmetry during the primordial epochs \cite{Rioto,Dine,Alexander/2006,Mohanty/2006,Li/2004,Lambiase/2013,Oikonomou/2016,Oikonomou/2017,Odintsov/2016}.
In this intricate tapestry of theoretical physics, researchers fervently seek to decipher the cosmic puzzle that underlies the baryon asymmetry, propelling our understanding of the fundamental nature of the Universe to new frontiers.\\
In the year 1967, Andrei Sakharov put forth a seminal proposition encompassing three indispensable prerequisites for the emergence of a baryon asymmetry,  Baryon-Number (Non-Conserving Interactions), C and CP violation and out of thermal equilibrium \cite{Sakharov/1967}. Sakharov's visionary scenario involves a Universe that initially contracts, possessing a baryon asymmetry of equal but opposite magnitude to that presently existing. This Universe then undergoes a pivotal moment, characterized by a bounce at the singularity, effectively reversing the magnitude of its baryon asymmetry. Recent research endeavors have unveiled the promising potential to instigate a baryon asymmetry by harnessing a subset of the three crucial conditions postulated by Sakharov. In a notable development, Cohen et al. \cite{Col/1993} have put forth an ingenious model for the creation of baryon number asymmetry. This model ingeniously hinges upon the violation of CPT (charge-parity-time) symmetry while concurrently maintaining thermal equilibrium. Such groundbreaking advancements in our comprehension of baryogenesis continue to reshape the frontiers of theoretical physics, pushing the boundaries of what we once deemed possible.\\
Recently, authors have explored the topic of baryogenesis within modified theories of gravity. The specific context of baryogenesis in f(R) theories of gravity has also been investigated by \cite{GG}. In \cite{ft}, the authors have examined gravitational baryogenesis within the framework of $f(T)$ gravities, where T represents the torsion scalar. Similarly, the question of gravitational baryogenesis in the context of Gauss-Bonnet braneworld cosmology has been investigated in \cite{fv}. Authors in \cite{Sahoo/2020}, have studied gravitational baryogenesis in nonminimally coupled $f(R,\mathcal{T})$ theories, $\mathcal{T}$ being the trace of energy-momentum tensor. And recently the authors in \cite{Swagat}, have studied gravitational baryogenesis in $f(T,\mathcal{T})$ gravity.\\

The present study aims to explore a generalized cosmological $f(Q)$ model to probe gravitational baryogenesis. The article is organized in the following manner. We start with the mathematical aspects of $f(Q)$ theory in the next section. Then the third section is devoted to investigate gravitational baryogenesis scenarios in the framework of $f(Q)$ gravity by assuming two models. Finally, in the last section we conclude our results.

\section{The Mathematical Formulation of $\boldsymbol{F(Q)}$ Theory}\label{sec2}

In $ f(Q)$ gravity theory, we consider the teleparallel torsion-free geometrical background that can be achieved by setting Riemannian tensor and torsion tensor to be zero i.e.  $R^\alpha_{\: \beta\mu\nu} = 0$ and $T^\alpha_{\ \mu\nu} = 0$. These constraints of symmetric teleparallelism offer a total inertial affine connection. In an arbitrary gauge, the general affine connection $X^\alpha_{\ \mu\nu}$ characterizing tensors in parallel transport and covariant derivative, can be parameterized in the following manner \cite{jimenez/2020},
\begin{equation}\label{2a}
X^\alpha \,_{\mu \beta} = \frac{\partial x^\alpha}{\partial \xi^\rho} \partial_ \mu \partial_\beta \xi^\rho
\end{equation}
However, for some specific gauge choice, this generic connection $X^\alpha \,_{\mu \beta}$ vanishes. This specific gauge is well known as the coincident gauge. In the $f(Q)$ class of theories, the central object is the non-metricity tensor arising due to the incompatibility of the connection of the theory. The non-metricity tensor corresponds to generic connection \eqref{2a} is defined as,
\begin{equation}\label{2b}
Q_{\alpha\mu\nu}\equiv\nabla_\alpha g_{\mu\nu}
\end{equation} 
In addition, we define the disformation tensor $L^\alpha_{\ \mu\nu}$ as the difference between the generic connection $X^\alpha_{\ \mu\nu}$ and the Levi-Civita one $\Gamma^\alpha_{\ \mu\nu}$ i.e. $L^\alpha_{\ \mu\nu}=X^\alpha_{\ \mu\nu}-\Gamma^\alpha_{\ \mu\nu}$. It follows that
\begin{equation}\label{2c}
L^\alpha_{\ \mu\nu}\equiv\frac{1}{2}(Q^{\alpha}_{\ \mu\nu}-Q_{\mu \ \nu}^{\ \alpha}-Q_{\nu \ \mu}^{\ \alpha})
\end{equation}
Now, we introduce the non-metricity scalar which is the key ingredient to describe the gravitation in $f(Q)$ theory, 

\begin{equation}\label{2d}
Q = -Q_{\lambda\mu\nu}P^{\lambda\mu\nu} 
\end{equation}
where,
\begin{equation}\label{2e}
4P^\lambda\:_{\mu\nu} = -Q^\lambda\:_{\mu\nu} + 2Q_{(\mu}\:^\lambda\:_{\nu)} + (Q^\lambda - \tilde{Q}^\lambda) g_{\mu\nu} - \delta^\lambda_{(\mu}Q_{\nu)}
\end{equation}
is the non-metricity conjugate term with $Q_\alpha = Q_\alpha\:^\mu\:_\mu \: \text{and}\:  \tilde{Q}_\alpha = Q^\mu\:_{\alpha\mu}$ being two traces of non-metricity tensor.
The $f(Q)$ gravity action in the symmetric teleparallelism background reads  \cite{JLL}
\begin{equation}\label{2f}
S= \int{\frac{1}{2}f(Q)\sqrt{-g}d^4x} + \int{L_m\sqrt{-g}d^4x}
\end{equation}
where, $f(Q)$ represents an arbitrary function of the scalar term $Q$, $L_m$ is the matter Lagrangian density, and $g=det( g_{\mu\nu} )$.
The variation of the action term \eqref{2f}, with respect to the metric, corresponds to the following field equation,

\begin{multline}\label{2g}
\frac{2}{\sqrt{-g}}\nabla_\lambda (\sqrt{-g}f_Q P^\lambda\:_{\mu\nu}) + \frac{1}{2}g_{\mu\nu}f+\\
f_Q(P_{\mu\lambda\beta}Q_\nu\:^{\lambda\beta} - 2Q_{\lambda\beta\mu}P^{\lambda\beta}\:_\nu) = -{T}_{\mu\nu}
\end{multline}

Here, ${T}_{\mu\nu}$ is the stress-energy tensor defined as,
\begin{equation}\label{2h}
{T}_{\mu\nu} = \frac{-2}{\sqrt{-g}} \frac{\delta(\sqrt{-g}L_m)}{\delta g^{\mu\nu}}
\end{equation}
Furthermore, when the hypermomentum is not present, the variation of equation \eqref{2f} with respect to the connection leads to the following connection field equation,
\begin{equation}\label{2i}
\nabla_\mu \nabla_\nu (\sqrt{-g}f_Q P^{\mu\nu}\:_\lambda) =  0 
\end{equation}

We begin with the following homogeneous and isotropic flat FLRW line element given in Cartesian coordinates,
\begin{equation}\label{3a}
ds^2= -dt^2 + a^2(t)[dx^2+dy^2+dz^2]
\end{equation}
The gauge considered in the line element \eqref{3a} is a coincident gauge coordinate, and thus it follows that the metric is only a fundamental variable. However, the generic connection in a different gauge choice leads to a non-trivial contribution to the field equations \cite{rs1,FDA}. The non-metricity scalar $Q$ for the metric \eqref{3a} is given as
\begin{equation}\label{3b}
 Q= 6H^2  
\end{equation}

The matter is considered to be a perfect fluid, and the energy-momentum tensor $T_{\mu\nu}$ for this matter is defined as follows
\begin{equation}
\label{9}
T_{\mu\nu}=(p+\rho)u_{\mu}u_{\nu}+pg_{\mu\nu},
\end{equation}
where $p$ and $\rho$ are the pressure and energy density of a perfect fluid, respectively, and $u_{\mu}$ is a four-velocity vector. Now, the Friedmann equations in GR-like form for an arbitrary $f(Q)$ function are given as,
\begin{equation}\label{3e}
Q f_{Q} +\rho = \frac{f}{2}  
\end{equation}
and 
\begin{equation}\label{3f}
4H f_{QQ}+(4\dot{H}+ 2Q) f_{Q}  =  2p + f 
\end{equation}

The above motion equations will be used for our further calculations in the upcoming section.

\section{Baryogenesis in $\boldsymbol{f(Q)}$ Gravity} \label{sec3}
In the context of $f(Q)$ gravity, we contemplate an interaction term that violates CP symmetry, and this term is produced by the baryonic asymmetry, taking a specific form
\begin{equation}
\label{11}
    \frac{1}{M_*^2} \int \sqrt{-g} dx^4 (\partial_\mu(-Q))J^\mu.
\end{equation}
where $M_*$ is the cut-off scale of the effective theory. The motivation behind the above interaction term is from the proposed $CP$-violating interaction by Davoudiasl et al. \cite{Davoudiasl/2004}. We have considered the non-metricity scalar as an alternative to the Ricci scalar in this interaction.  
We proceed with the assumption that the reason behind the asymmetry is not following Sakahrov's third condition. That is thermal equilibrium is maintained in the Universe, with the energy density being directly proportional to the decoupling temperature $T_D^*$ as
\begin{equation}
\label{12}
    \rho = \frac{\pi^2}{30} g_* (T_D^*)^4,
\end{equation}
Here, the symbol  $g_*= \frac{45s}{2 \pi^2 (T_D^*)^3}$ signifies the number of degrees of freedom for the particles that contribute to the overall entropy of the Universe.  To get a measurement of the imbalance between matter and antimatter, baryon number density$(n_B)$ to entropy$(s)$ ratio$(n_B/s)$ is required. Various observations and predictions have confirmed the asymmetry with the $n_B/s$ value. BBN observation found $n_B/s=(5.6 \pm 0.6) \times 10^{-10}$ and Cosmic Microwave Background predicted $n_B/s=(6.19 \pm 0.14) \times 10^{-10}$ at $95 \%$ Confidence Level. The numerically observed baryon-to-entropy ratio value is $n_{B}/s=9.2^{+0.6}_{-0.4} \times 10^{-11}$. \\
We define the baryon-to-entropy ratio for the $CP$-Violating interaction term\eqref{11} as
\begin{equation}
\label{13}
    \frac{n_B}{S} \simeq -\frac{15g_b}{4\pi^2 g_*}\left[\frac{1}{M_*^2 T^*}(\dot{Q} )\right]_{T^*=T_D^*}, 
\end{equation}
with $g_b$ being the total number of intrinsic degrees of freedom of baryons.\\
\subsection{Power Law Model}

We consider the nonlinear polynomial form of $f(Q)$ as
\begin{equation}
\label{14}
    f(Q)= \alpha Q^n 
\end{equation}
where $n$ and $\alpha$ are arbitrary parameters. It is obvious that the model reduces to GR with the limit $\alpha=n=1$.\\

For this model, one can find the scale factor by inserting \eqref{14} in the first motion equation \eqref{3e}  as $a(t)=(B t)^{2n/3}$ where  $ B=\frac{3}{2n}\left(\frac{2{\rho}_0}{\alpha6^n(2n-1)}\right)^{\frac{1}{2n}}$, $\rho_0$ is the present density.
Similarly, the expression for energy density can be found as
\begin{equation}
\label{16}
    \rho=\alpha\left(n-\frac{1}{2}\right)\left(\frac{8n^2}{3t^2}\right)^n 
\end{equation}
The obtained energy density can now be equated to the assumed one \eqref{12} to obtain the decoupling time $(t_D)$ in terms of the decoupling temperature $T_D^*$. By doing so we find,
\begin{equation}
\label{x}
    t_D=\left(\frac{30 \alpha \left(n-\frac{1}{2}\right) (8 n^2)^n}{3^n \pi^2 g_* {T_D^*}^4}\right)^\frac{1}{2n}
\end{equation}

Now, the model parameter $\alpha$ can be obtained in terms of $n$ at present time from the first motion equation as,

\begin{equation}
\label{17}
    \alpha= \frac{\Omega_{m_0}}{6^{n-1} H_0^{2n-2}(2n-1)}
\end{equation}
where $\Omega_{m_0}$ and $H_0$ represent the density and Hubble parameter for the present time respectively.\\
Using the decoupling time along with $\alpha$ in the baryon to entropy ratio \eqref{13} we obtain,
\begin{multline}
    \label{18}
    \frac{n_B}{s} \simeq -\frac{15 g_b}{4 \pi^2 g_* M_*^2 T_D^*}
    \left(\frac{-16}{3} 5^{-\frac{3}{2 n}} n^2 \pi ^{3/n}\right) \\ \times \left(\frac{2^{3 n+1} 3^{1-n} \left(n-\frac{1}{2}\right) \left(n^2\right)^n \Omega _{m_0}}{g_* {T_D^*}^4 (2n-1)\left(6^{n-1} H_0^{2n-2}\right)}\right)^{-\frac{3}{2 n}}
\end{multline}

\begin{figure}
    \centering
    \includegraphics[width=1\linewidth]{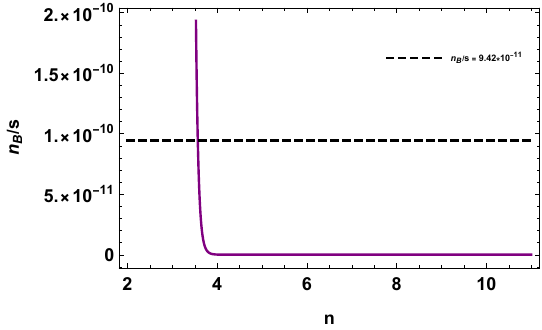}
    \caption{The model parameter $n$ vs baryon-to-entropy ratio for the Power law model }
     \label{fig 1}
\end{figure}

The obtained baryon-to-entropy ratio \eqref{18} for this model has many parameters with prior values from various experiments and observations. So our objective narrows down to constrain the model parameter $n$ to match with the observed value of ${n_B}/s$.  From Fig. \ref{fig 1} we depict that for $n= 3.569$, $n_B/s$ agrees with the observational value $9.42\times 10^{-11}$ (represented by the black dashed line). The other model parameter $\alpha$ can be obtained from \eqref{17} by fixing $\Omega_{m_0}=0.3$ and $H_0=67.2$ \cite{{Aghanim/2020},{Mandal/2023}}  as, $\alpha=1.99 \times 10^{-13}$. The remaining parameters are fixed with the values $g_b=1$, $T_D^*=2\times10^{16} GeV$,  $M_*=10^{12} GeV$ and $g_*=106$. 

\subsection{DGP-like $f(Q)$ model}
Now we consider the recently introduced \cite{Ayuso/2022} DGP-like $f(Q)$ model,
\begin{equation}
\label{19}
    f(Q)= \beta_1 \sqrt{Q} log Q +2 \beta_2 Q
\end{equation}
where $\beta_1$ and $\beta_2$ are the free parameters. The motivation behind this model is the requirement of an additional term in the modified Friedmann equation which is equivalent to $H$. So, considering $\sqrt{Q} $ in the model will come up with an extra $H$ term on the left-hand side of the Friedmann equation which makes it eligible to explain several modified gravity scenarios. This model simply reduces to GR with the corresponding values of model parameters as $\beta_1=0$ and $\beta_2=1/2$.\\
Further, we start by assuming a power-law cosmic evolution characterized by the scale factor as $a(t)=At^r$. Using $a(t)$ along with the lagrangian \eqref{19} in the first Friedmann equation \eqref{3e}, the analytic expression for the energy density can be found as
\begin{equation}
\label{20}
    \rho=6 \beta_2 \frac{r^2}{t^2} + \sqrt{6} \beta_1 \frac{r}{t}
\end{equation}
Now equating \eqref{20} with \eqref{12}, we get the explicit form of decoupling time in terms of decoupling temperature as,
\begin{equation}
\label{21}
    t_D=\frac{\sqrt{6} \beta_1 + \sqrt{6 {\beta_1}^2 + \frac{4}{5} \beta_2 r^2 \pi^2 g_* {T_D^*}^4}}{\frac{\pi^2}{15} g_* {T_D^*}^4}
\end{equation}
Similarly to the polynomial model, one of the model parameters can be obtained in terms of the other at present as,
\begin{equation}
\label{22}
   \beta_2 =\frac{1}{2}\left( \Omega_{m_0}-\frac{\sqrt{6} \beta_1}{3 H_0)}\right)
\end{equation}

Now we insert the decoupling time \eqref{21} and $\beta_2$ in \eqref{13} to obtain the baryon-to-entropy ratio for this model. We find,

\begin{equation} \label{23}
 \frac{n_B}{s} \simeq \frac{\pi ^4 \,\ g_b \,\ g_*^2 \,\ r^2  {T_D^*}^{11}}{75 M_*^2 \left(\sqrt{6} \beta_1 +\sqrt{C}\right)^3} 
\end{equation}

where $C=6 \beta_1 ^2+\frac{2}{5} \pi ^2 g_* r^2 {T_D^*}^{4} \left(\Omega_{m_0} -\frac{\sqrt{\frac{2}{3}} \beta_1 }{H_0}\right)$.

\begin{widetext}

\begin{figure}[H]
     \centering

    \includegraphics[width=2.25\linewidth]{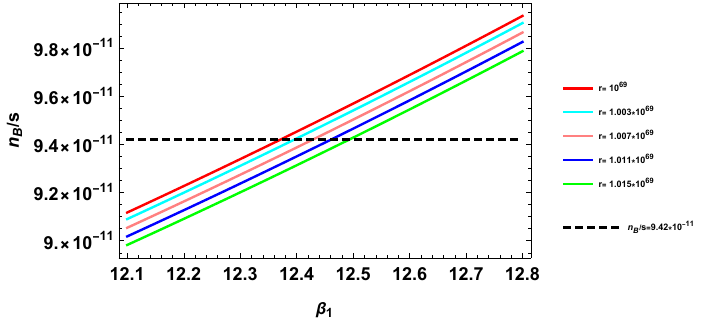}
    \caption{The model parameter $\beta_1$ vs baryon-to-entropy ratio for five different values of $r$.}
     \label{fig 2}
\end{figure}
\end{widetext}

From Fig.\ref{fig 2} we depict that the model agrees with the numerically observed value of the baryon-to-entropy ratio for various combinations of $\beta_1$ and $r$. We constrain them for five different values of $r$ which are represented by the colored curves in the plot. Also, the black dashed line represents the value $n_B/s= 9.42 \times 10^{-11}$. The other model parameter values along with the exact baryon-to-entropy ratio value for each case can be found in Table \ref{table1}. The remaining parameters are taken as $\Omega_{m_0}=0.3$, $H_0=67.2$, 
$g_b=1$, $T_D^*=2\times10^{16} GeV$,  $M_*=10^{12} GeV$ and $g_*=106$.

\begin{table}[H]
 \centering
 \caption{ Model parameters and $n_B/s$ values for the five
considered cases in the DGP-like $f(Q)$ model.
 }
 
 \label{table1}
    \begin{tabular}{||c||c|c|c||}
    \hline
    $r$ & $\beta_1$ & $\beta_2$ & $n_B/s$ \\
    \hline
     $10^{69}$ & $12.38$ & $0.07479$ & $9.42956 \times 10^{-11}$ \\
    
   $1.003 \times 10^{69}$  & $12.40$ & $0.0746685$ & $9.42431 \times 10^{-11}$ \\
    
    $1.007 \times 10^{69}$ & $12.43$ & $0.0744862$ & $9.42135 \times 10^{-11}$ \\

    $1.011 \times 10^{69}$ & $12.467$ & $0.0742614$ & $9.42671 \times 10^{-11}$ \\

    $1.015 \times 10^{69}$ & $12.5$ & $0.074061$ & $9.42772 \times 10^{-11}$ \\
    \hline

    \end{tabular}
\end{table}


\section{Conclusion}\label{sec4}
\justify This work was devoted to study the gravitational baryogenesis scenario in the framework of $f(Q)$ gravity. A $CP$-violating interaction term that satisfies the second condition of Sakharov has been considered, which could be the possible reason for the baryon asymmetry. The corresponding baryon-to-entropy ratio is defined in the context of non-metricity scalar. We take the nonlinear polynomial model $  f(Q)= \alpha Q^n$ as our first model and find that it completely agrees with the numerically observed baryon-to-entropy ratio for the model parameter value $n=3.569$. The second model parameter, which was eliminated in terms of $n$ at the present time, can be found as $\alpha=1.99 \times 10^{-13}$.  Further, we take the recently proposed DGP-like model $f(Q)= \beta_1 \sqrt{Q} log Q +2 \beta_2 Q $. We notice that it is readily attainable to obtain a baryon-to-entropy ratio that aligns with the observed value $\frac{n_b}{s}=9.42 \times 10^{-11}$. We constrain five different values of $r$ with the model parameter $\beta_1$. In all the computations, we have considered $g_b=1$, $T_D^*=2\times10^{16} GeV$, $M_*=10^{12} GeV$ and $g_*=106$. Finally, we conclude that both our models suitably describe the gravitational baryogenesis with some constraints on the model parameters.

\acknowledgments
SSM acknowledges the Council of Scientific and Industrial Research (CSIR), Government of India for awarding Junior Research fellowship (E-Certificate No.: JUN21C05815). Aaqid Bhat expresses gratitude to the BITS-Pilani, Hyderabad Campus, India, for granting him a Junior Research Fellowship. PKS acknowledges Science and Engineering Research Board, Department of Science and Technology, Government of India for financial support to carry out Research project No.: CRG/2022/001847 and IUCAA, Pune, India for providing support through the visiting Associateship program. We are very much grateful to the honorable referee and to the editor for the illuminating suggestions that have significantly improved our work in terms of research quality, and presentation.\\

\textit{Data availability statement}: No new data were created or analysed in this study.

\end{document}